\documentclass[twocolumn,showpacs,preprintnumbers,amsmath,amssymb]{revtex4}
\usepackage{graphicx,epsfig}
\usepackage{dcolumn}
\usepackage{bm}
\newcommand{\beq}{\begin{equation}}
\newcommand{\eeq}{\end{equation}}
\newcommand{\beaq}{\begin{eqnarray}}
\newcommand{\eeaq}{\end{eqnarray}}
\newcommand{\bwt}{\begin{widetext}}
\newcommand{\ewt}{\end{widetext}}
\newfam\msbfam
\font\twlmsb=msbm10 at 12pt
\font\eightmsb=msbm10 at 8pt
\font\sixmsb=msbm10 at 6pt
\textfont\msbfam=\twlmsb
\scriptfont\msbfam=\eightmsb
\scriptscriptfont\msbfam=\sixmsb
\def\cj{\fam\msbfam}

\def\C{{\cj C}}

\def\R{{\cj R}}

\newcommand{\dt}{\delta t}
\newcommand{\dtt}{(\delta t)^2}
\begin{document}
\title{Quantum Mechanics and the Weak Equivalence Principle}

\author{Stella Huerfano$^{1,2}$, Sarira Sahu$^2$ and M. Socolovsky$^2$}

\affiliation{
$^1$Departamento de Matem\'aticas, Universidad Nacional de 
Colombia, Bogot\'a, Colombia\\
$^2$ Instituto de Ciencias Nucleares, 
Universidad Nacional Aut\'onoma de 
M\'exico\\
Circuito Exterior, Ciudad Universitaria, \\
04510, M\'exico D. F., M\'exico}
\thispagestyle{empty}

\begin{abstract}

\noindent We use the Feynman path integral approach to nonrelativistic quantum 
mechanics twofold. First, we derive the lagrangian for a spinless particle 
moving in a uniformly but not necessarily constantly accelerated reference 
frame; then, applying the {\it strong equivalence principle} (SEP) we obtain 
the Schroedinger equation for a particle in an inertial 
frame and in the presence of a uniform and constant gravity field. Second, 
using the associated Feynman propagator, we propagate an initial gaussian wave 
packet, with the final wave function and probability density depending on the 
ratio $\frac{m}{\hbar}$, where $m$ is the inertial mass of the particle, 
thus exhibiting the fact that {\it the weak equivalence principle} (WEP) {\it 
is violated by quantum mechanics}. Although due to rapid oscillations the wave 
function does not exist in the classical limit, the probability density is 
well defined and mass independent when $\hbar \to 0$, showing the recovery of 
the WEP. Finally, at the quantum level, a heavier particle does not 
necessarily 
falls faster than a lighter one; this depends on the relations between the 
initial and final common positions and times of the particles.
\end{abstract}
\maketitle
\footnotetext[1]{e-mails: rshuerfanob@unal.edu.co, 
sarira@nucleares.unam.mx, 
socolovs@nucleares.unam.mx}

\section{Introduction}

The {\it strong equivalence principle} (SEP) says that a reference frame 
accelerated with respect to an inertial system is equivalent to a uniform 
gravitational field in an inertial frame; and the other way around, an 
arbitrary gravitational field in an inertial system is locally equivalent to 
an accelerated frame. The {\it weak equivalence principle} (WEP), on  the 
other hand, says that under identical initial conditions, the motion of 
particles in a given gravitational field is the same; in particular, it is 
independent of their inertial or gravitational masses\cite{weinberg}. In this
context, it is important to emphasize the following points:

\noindent i) Both the SEP and the WEP are classical i.e. not quantum.

\noindent ii) The SEP implies the WEP: in fact, the accelerations of two 
particles with inertial masses $m_1$ and $m_2$ in a non inertial frame 
(and therefore in an equivalent gravitational field) are the same and 
independent of their masses; so their motions are equal.

\noindent iii) In the context of Newtonian mechanics it can be easily shown 
that the WEP is equivalent to the statement of the equality of the inertial 
mass $m$ and the (passive) gravitational mass $m_{g}$\cite{couderc} 
(see section 2).

\noindent iv) Even in the context of special relativity, it can be 
shown\cite{landau,greenberger} that in accelerated reference frames, 
the space (spacetime) 
geometry is not euclidean (not pseudoeuclidean) but riemannian
(pseudoriemannian); this 
fact supports the geometrical Einstein's theory of gravitation i.e. general 
relativity\cite{einstein}.

As is well known, in nonrelativistic quantum mechanics (NRQM), the motion of 
a particle in the presence of an external gravitational field is mass 
dependent\cite{sakurai,colella}; this can be seen from the Schroedinger 
equation in the 
local gravitational potential $V=gx$. Once the SEP is accepted one obtains 
\beq
i\hbar {{\partial}\over{\partial t}}\psi(x,t)
=(-{{\hbar ^2}\over{2m}}{{\partial ^2}\over{\partial x^2}}+mgx)\psi(x,t),
\eeq 
i.e.
\beq
i{{\partial}\over{\partial t}}\psi(x,t)
=(-{{\hbar}\over{2m}}{{\partial ^2}\over{\partial x^2}}
+{{m}\over{\hbar}}gx)\psi(x,t), 
\label{scheq}
\eeq
whose solution depends parametrically on ${{\hbar}\over{m}}$. 
Moreover, even the free motion is mass dependent, not only in NRQM but also 
in relativistic quantum mechanics (RQM):
\beq
(\partial ^2+\lambda_c^{-2})\varphi(x^{\mu})=0, 
\  \  \ Klein-Gordon \  \  \ equation,
\eeq
and
\beq
(i\gamma\cdot \partial+\lambda_c^{-1})\psi(x^\mu)=0, 
\  \  \  Dirac \  \  \ equation,
\eeq
where $\lambda_c$ is the Compton length given by ${{\hbar}\over {mc}}$. 
In NRQM the free propagator of a spinless particle of mass $m$ is given by 
\beq
K_0(x^{\prime\prime},t^{\prime\prime};x^\prime,t^\prime)
=\sqrt{{{m}\over{2\pi i \hbar(t^{\prime\prime}-t^\prime)}}}
e^{{{im}
\over{\hbar}}{{(x^{\prime\prime}-x^\prime)^2}
\over{t^{\prime\prime}-t^\prime}}}, 
\eeq
that replaces the mass independent solution of the corresponding free 
motion in classical mechanics:

\newpage

\beq
velocity= const.
={{x^{\prime\prime}-x^\prime}\over{t^{\prime\prime}-t^\prime}}.
\eeq 
The above arguments show that in quantum mechanics, for wave functions, 
propagators and probability density distributions, {\it the WEP is violated}. 
However, this result does not imply neither the violation of the SEP, as 
claimed in Ref.\cite{rabinowitz}, since the implication 
$SEP \Longrightarrow WEP$ is 
classical but not quantum, nor the violation of the WEP, at the level of 
expectation values, where it indeed holds\cite{viola}. 
Though the wave function and 
the propagator are not observables, but the position probability density 
distribution can be measured, then {\it the violation of the WEP in 
quantum mechanics is physical}. We do not discuss here the generalization 
of equivalence principles to 
quantum mechanics (see e.g. Ref.\cite{lammerzhal}).

In sec.2, assuming the SEP, we rederive the Eq.(\ref{scheq}) 
in the context 
of the path integral formulation of quantum mechanics\cite{feynman}, which 
involves the classical lagrangian. We also show the equivalence between the 
equality $m=m_{g}$ (up to a universal constant) and the WEP. 

In sec.3, we use the Feynman propagator in the presence of a uniform and 
constant gravitational field $\vec{g}$, and study the quantum free fall of a 
spinless particle of mass $m$ that at time $t^\prime$ has a gaussian 
distribution of width $\sigma$ around $x^\prime$ and average momentum $p_0$; 
the probability density at time $t^{\prime\prime}>t^\prime$ is also gaussian,
 but mass dependent. We also discuss the difference in the final probability 
density distributions $\rho_1$ and $\rho_2$ corresponding to two 
different particles of 
masses $m_1$ and $m_2$, with $m_1>m_2$. A similar analysis but in the 
framework of the causal interpretation of quantum mechanics 
(Bohmian mechanics) was done in Ref.\cite{holland}.

Finally, in sec.4 the classical limit $\hbar \to 0$ is taken and the 
mass dependence of quantum free fall which is discussed in sec. 3 
disappears, as it should be to recover the 
classical WEP. The limit $\sigma \to 0$ for the ideal case of perfectly 
localized particles is shown to behave in the same way.

\section{Uniform acceleration and SEP in non-relativistic
 quantum mechanics}

Let $t^{\prime\prime}>t^\prime$; then, in an inertial reference frame with 
coordinates $(\vec{x},t)$ the Feynman propagator between the points 
$(\vec{x}^\prime,t^\prime)$ and $(\vec{x}^{\prime\prime},t^{\prime\prime})$ 
of a non-relativistic particle with inertial mass $m$ in a time independent 
potential $U$ can be represented by an integral over all continuous paths 
joining the above initial and final points\cite{feynman}:
\bwt 
\beq
K(\vec{x}^{\prime\prime},t^{\prime\prime};\vec{x}^\prime,t^\prime)
=\int _{\vec{y}(t^\prime)=\vec{x}^\prime}^{\vec{y}(t^{\prime\prime})
=\vec{x}^{\prime\prime}}{\cal D}\vec{y}(t)~
exp \left [ \frac{i}{\hbar} \int dt 
\left ( \frac{1}{2} m\vert\dot{\vec{y}}(t)\vert^2 - U(\vec{y}-\vec{y}_0) 
\right )\right ].
\label{kernel1}
\eeq
\ewt
$\vec{y}_0$ is an arbitrary reference point. Going  to a reference frame with 
spacetime coordinates 
\beq
\vec{\tilde{y}}=\vec{y}-\vec{\xi}(t)
\eeq
and 
\beq
\tilde{t}=t 
\eeq
where $\vec{\xi}(t)$ is an arbitrary  twice differentiable function of 
the time, after integration, Eq.(\ref{kernel1}) becomes 
\bwt
\beaq
&&K(\vec{x}^{\prime\prime}, t^{\prime \prime};\vec{x}^\prime,t^\prime)
= exp\left [
\frac{im}{\hbar} \left (\vec{\tilde{x}}^{\prime\prime}
\cdot \dot{\vec{\xi}}(t^{\prime\prime})+\frac{1}{2}
\int_{t^\prime}^{t^{\prime\prime}}dt\vert\dot{\vec{\xi}}(t)\vert^2
\right)\right ]
\nonumber\\
&&\times\int_{\vec{\tilde{y}}(t^\prime)
=\vec{\tilde{x}}^\prime}^{\vec{\tilde{y}}(t^{\prime\prime})
=\vec{\tilde{x}}^{\prime\prime}}{\cal D}\vec{\tilde{y}}~
exp\left [
\frac{i}{\hbar}
\int_{t^\prime}^{t^{\prime\prime}}dt \left ({{m}\over {2}}
\vert \dot{\vec{\tilde{y}}}\vert^2-(m\vec{\tilde{y}}
\cdot \ddot {\vec{\xi}}+U(\vec{\tilde{y}}-\vec{\tilde{y}}_0))\right )
\right ]
e^{-{{im}\over{\hbar}}\vec{\tilde{x}}^\prime
\cdot \dot{\vec{\xi}}(t^\prime)},
\eeaq
\ewt
where 
$\vec{\tilde{x}}^\prime=\vec{x}^\prime-\vec{\xi}(t^\prime)$, 
$\vec{\tilde{x}}^{\prime\prime}
=\vec{x}^{\prime\prime}-\vec{\xi}(t^{\prime\prime})$ and 
${\cal D}\vec{y}={\cal D}\vec{\tilde{y}}$. 
Defining the wave function 
\beq
\tilde{\psi}(\vec{\tilde{x}}, \tilde{t})
=e^{-{{im}\over\ {\hbar}}(\vec{\tilde{x}}\cdot \dot{\vec{\xi}}(t)
+{{1}\over{2}}\int_{\bar{t}}^{\tilde{t}}d\tau 
\vert \dot {\vec{\xi}}(\tau)\vert ^2)}\psi(\vec{x},t)
\eeq
where $\bar{t}$ is an arbitrary instant between $t^\prime$ and 
$t^{\prime\prime}$, one obtains
\beq 
\tilde{\psi}(\vec{\tilde{x}}^{\prime\prime}, \tilde{t}^{\prime\prime})
=\int d\vec{\tilde{x}}^\prime \tilde{K}(\vec{\tilde{x}}^{\prime\prime},
\tilde{t}^{\prime\prime};\vec{\tilde{x}}^\prime,
t^\prime)\tilde{\psi}(\vec{\tilde{x}}^\prime,\tilde{t}^\prime)
\eeq
with
\beq 
\tilde{K}(\vec{\tilde{x}}^{\prime \prime},\tilde{t}^{\prime\prime};
\vec{\tilde{x}}^\prime,\tilde{t}^\prime)
=\int_{\vec{\tilde{x}}^\prime}^{\vec{\tilde{x}}^{\prime\prime}}{\cal D}
\vec{\tilde{y}}e^{{{i}\over{\hbar}}
\int_{\tilde{t}^\prime}^{\tilde{t}^{\prime\prime}}dt({{m}\over{2}}
\vert\dot{\vec{\tilde{y}}}(t)\vert^2-U_{eff})}
\eeq
and where $U_{eff}$ is an effective potential naturally incorporating the 
effect of the acceleration $\ddot{\vec{\xi}}(t)$: 
\beq
U_{eff}=m\vec{\tilde{y}}\cdot\ddot{\vec{\xi}}+U(\vec{\tilde{y}}
-\vec{\tilde{y}}_0). 
\eeq 
The SEP allows us to reinterpret 
$\tilde{K}$ as the Feynman propagator in an inertial frame in the presence 
of a uniform but otherwise arbitrary gravitational field given by 
$\ddot{\vec{\xi}}(t)$\cite{weinberg}. In particular, for constant 
\beq
\ddot{\vec{\xi}}=\vec{g}
\eeq
we obtain the usual Feynman  propagator for a spinless quantum particle 
coupled to a constant and uniform gravitational field $\vec{g}$:
\bwt 
\beq
K(\vec{x}^{\prime\prime},t^{\prime\prime};\vec{x}^\prime,t^\prime;\vec{g};
{{m}\over{\hbar}})
=\int_{\vec{y}(t^\prime)=\vec{x}^\prime}^{\vec{y}(t^{\prime\prime})
=\vec{x}^{\prime\prime}}{\cal D}\vec{y}~exp\left [({{i}
\over{\hbar}}\int_{t^\prime}^{t^{\prime\prime}}dt
\left (
{{m}\over{2}}
\vert\dot{\vec{y}}(t)\vert^2-(m\vec{y}(t)\cdot\vec{g}
+U(\vec{y}-\vec{y}_0))\right )\right ], 
\label{kernel2}
\eeq
\ewt 
where we emphasized the dependence of $K$ on $\vec{g}$ and on the ratio 
${{m}\over{\hbar}}$.

Notice that the assumption of the validity of the SEP also in quantum 
mechanics, has implied the equality of the inertial mass $m$ with the 
{\it passive gravitational mass} $m_{g}$, which gives the coupling between 
the particle and the gravitational field: 
\beq
U_{g}=m_{g}\vec{\tilde{y}}\cdot \vec{g}. 
\eeq 
Classically, the equality $m=m_{gr}$ (or $m=km_{gr}$ with $k$ a 
universal constant) is {\it equivalent} to the WEP, which says that under 
identical initial conditions the 
motion (acceleration) of particles in a given gravitational field is 
independent of their masses\cite{couderc}. In fact, for two particles 
with inertial 
masses $m$ and $M$ and corresponding passive gravitational masses $m_{g}$ 
and $M_{g}$, the Newton equations are $a={{m_{g}}\over{m}}g$ and 
$A={{M_{g}}\over{M}}g$; then the WEP implies $a=A=\tilde{g}$ and therefore 
$\nu_{g}=k\nu$ for both $\nu=m$ and $\nu=M$ ($\tilde{g}=kg$ and 
$\tilde{g}=g$ only if the units are chosen such that $k=1$); the other way 
around: if $\nu_{g}=k\nu$ then $\nu a=\nu_{g}g=k\nu g$ and then $a=kg$ 
for both $m$ and $M$. So, classically, 
\beq
(m_{g}=km) \buildrel \ {cl}\over \Longleftrightarrow WEP. 
\eeq
But also we have that both in quantum mechanics as well as classically, 
\beq
SEP\buildrel {QM/cl}\over \Longrightarrow (m_{g}=km). 
\eeq
Then we have the chain of implications 
\beq
SEP \buildrel {QM/cl}\over \Longrightarrow (m_{g}=km) \buildrel {cl} 
\over \Longleftrightarrow WEP.
\eeq

The Schroedinger equation which emerges from Eq.(\ref{kernel2}) 
is\cite{feynman}
\beq 
i{{\partial}\over{\partial t}}\psi(\vec{y},t)
=(-{{1}\over{2}}{{\hbar}\over{m}}\nabla^2+({{m}\over{\hbar}}),
\vec{y}\cdot \vec{g})\psi(\vec{y},t) 
\eeq
where we have set $U(\vec{y}-\vec{y}_0)=0$. This formula has been 
experimentally verified by the now famous COW experiment\cite{colella} 
using neutron 
interferometry. Both the Schroedinger equation as well as the propagator 
$\tilde{K}$ are mass dependent which strongly suggests that the WEP is 
violated by quantum mechanics. But this does not invalidate neither the 
equality $m_{g}=km$ (since the implication 
$(m_{g}=km)\buildrel {cl}\over \Longrightarrow WEP$ involves the classical 
Newton equation) nor the SEP which remains true both in classical mechanics 
as well as in quantum mechanics. Obviously, in $QM$ we have the result that 
\beq
(m_{g}=km) \ \ is \ \ not \ \ equivalent \ \ to \ \ \ WEP. 
\eeq
This is discussed in the book by Sakurai\cite{sakurai}.

\section{Propagation of a gaussian wave packet in the local gravitational 
field and violation of the WEP}

To study the free fall of a quantum particle of mass $m$ in a local 
gravitational potential $gx$, we need the propagator 
$K(x^{\prime\prime},t^{\prime\prime};x^\prime,t^\prime)$. 
For simplicity, we shall consider the whole vertical axis $x$ as the domain 
of the motion, ignoring the infinite barrier imposed by the 
surface of the earth.

For a quadratic lagrangian of the form 
\bwt
\beq
L(x,\dot{x},t)={{1}\over{2}}m\dot{x}(t)^2+b(t)x(t)
\dot{x}(t)+d(t)\dot{x}(t)-{{1}\over{2}}c(t)x(t)^2-e(t)x(t)-f(t)
\eeq 
\ewt 
the propagator is given by\cite{feynman,schulman}
\beq
K(x^{\prime\prime},t^{\prime\prime};x^\prime,t^\prime)
=\sqrt{{{m}\over{2\pi i \hbar f(t^{\prime\prime},t^\prime)}}}
e^{{{i}\over{\hbar}}S[\bar{x}]}
\eeq
where $\bar{x}(t)$ is the classical path joining the initial and final 
points $(x^\prime,t^\prime)$ and $(x^{\prime\prime},t^{\prime\prime})$, 
and $f(\xi,\eta)$ satisfies the differential equation 
\beq
{{\partial^2}\over {\partial \xi^2}}f(\xi,\eta)+{{\dot{b}(\xi)
+c(\xi)}\over{m}}f(\xi,\eta)=0,
\label{eqmotion}
\eeq
with the conditions 
\beq
f(\xi,\xi)=0, \ \ \ {{\partial}\over{\partial \xi}}
f(\xi,\eta)\vert_{\xi=\eta}=1. 
\eeq
In our case, from Eq.(\ref{kernel2}) with $U(\vec{y}-\vec{y}_0)=0$, the 
lagrangian becomes
\beq 
L={{m}\over{2}}\dot{x}^2-mgx(t),
\eeq
i.e. 
$b(t)=d(t)=c(t)=f(t)=0$ and $e(t)=mg$; then the Eq.(\ref{eqmotion}) reduces 
to ${{\partial^2}\over {\partial \xi^2}}f(\xi,\eta)=0$ with solution 
$f(\xi,\eta)=\xi\alpha(\eta)+\beta(\eta)$; from the initial conditions, 
$\eta \alpha(\eta)+\beta(\eta)=0$ which implies 
$\beta(\eta)=-\eta\alpha(\eta)$ and $\alpha(\eta)=1$, then 
$\beta(\eta)=-\eta$ and so $f(\xi,\eta)=\xi-\eta$. Then 
$f(t^{\prime\prime},t^\prime)=t^{\prime\prime}-t^\prime$ and the propagator 
becomes 
\beq
K(x^{\prime\prime},t^{\prime\prime};x^\prime,t^\prime)
=\sqrt{{{m}\over{2\pi i \hbar (t^{\prime\prime}-t^\prime)}}}e^{{{i}
\over{\hbar}}S[\bar{x}]}
\eeq
with 
\beq
\bar{x}(t)=x^\prime+v_0(t-t^\prime)-{{1}
\over{2}}g(t-t^\prime)^2, 
\eeq
where $v_0$ is the initial velocity given by 
\beq
v_0={{x^{\prime\prime}-x^\prime}\over{t^{\prime\prime}-t^\prime}}
+{{g}\over{2}}(t^{\prime\prime}-t^\prime). 
\eeq
For the classical action one has 
\bwt
\beq
S[\bar{x}]=
\int_{t^\prime}^{t^{\prime\prime}}dt
\left [
{{1}\over{2}}\dot{\bar{x}}(t)^2
-mg\bar{x}(t)
\right ]
={{m}\over{2}}
(t^{\prime\prime}-t^\prime)
\left [\left ({{x^{\prime\prime}
-x^\prime}\over{t^{\prime\prime}-t^\prime}}\right )^2
-g(x^{\prime\prime}+x^\prime)
-{{1}\over{12}}g^2(t^{\prime\prime}-t^\prime)^2
\right ]
\eeq
\ewt
which gives the propagator, explicitly depending on ${{m}\over{\hbar}}$,
\bwt
\beq 
K(x^{\prime\prime},t^{\prime\prime}-t^\prime,x^\prime;{{m}\over{\hbar}})
=\sqrt{{{m}\over{2\pi i \hbar (t^{\prime\prime}-t^\prime)}}}
e^{{{im}
\over{2\hbar}}(t^{\prime\prime}-t^\prime)
\left [\left ({{x^{\prime\prime}
-x^\prime}\over{t^{\prime\prime}-t^\prime}}\right )^2
-g(x^{\prime\prime}+x^\prime)-{{1}
\over{12}}g^2(t^{\prime\prime}-t^\prime)^2\right ]}. 
\eeq
\ewt
In the classical limit, $K$ does not exist by the rapid oscillations of the 
exponential; however, 
\beq
\vert K(x^{\prime\prime},t^{\prime\prime}
-t^\prime,x^\prime;{{m}\over{\hbar}})\vert \buildrel 
\ {\hbar \to 0}\over \longrightarrow +\infty.
\eeq 
We remark that even the free nonrelativistic propagator depends on $m$: 
\beq
K_0(x^{\prime\prime}-x^\prime,t^{\prime\prime}-t^\prime;
{{m}\over{\hbar}})=\sqrt{{{m}\over {2\pi i \hbar (t^{\prime\prime}
-t^\prime)}}}e^{{{im}\over{2\hbar}}{{(x^{\prime\prime}-x^\prime)^2}
\over{t^{\prime\prime}-t^\prime}}}
\eeq
which says that even the free motion is mass dependent. The same happens 
for the propagators in the relativistic domain, like those for the
 Klein-Gordon and Dirac particles. This is a first indication of the 
violation of the WEP in the quantum regime.

If $\psi(x^\prime,t^\prime)$ is the initial wave function describing our 
``falling" particle, then the wave function at 
$(x^{\prime\prime},t^{\prime\prime})$ is given by 
\beq
\psi(x^{\prime\prime},t^{\prime\prime})=\int_{-\infty}^{+\infty}dx^\prime
 K(x^{\prime\prime},t^{\prime\prime};x^\prime,t^\prime)\psi(x^\prime,t^\prime).
\eeq
For $\psi(x^\prime,t^\prime)$ we choose a normalized gaussian wave packet 
centered at $x^\prime$, average momentum $p_0=\hbar k_0$ and therefore 
average velocity $u_0={{p_0}\over{m}}={{\hbar k_0}\over{m}}$, and width 
$\sigma$, namely 
\beq
\psi(y,t^\prime)={{e^{-{{(y-x^\prime)^2}\over{2\sigma^2}}+ik_0y}}
\over{\pi^{{{1}\over{4}}}\sqrt{\sigma}}}.
\eeq
Then, the wave function at $(x^{\prime\prime},t^{\prime\prime})$ is 
\bwt
\beaq
&&\psi(x^{\prime\prime},t^{\prime\prime}-t^\prime,x^\prime; g;\sigma;{{m}
\over{\hbar}};k_0)
=\int_{-\infty}^{+\infty} dy K(x^{\prime\prime},
t^{\prime\prime};y,t^\prime)\psi(y,t^\prime)
\nonumber\\
&&
=\sqrt{{{m}\over{2\pi^{{3}\over{2}}\hbar i \sigma (t^{\prime\prime}
-t^\prime)}}}
e^{{{im}\over{2\hbar}}
\left [
{{x^{\prime\prime}}\over{(t^{\prime\prime}
-t^\prime)}}\left (x^{\prime\prime}-g(t^{\prime\prime}-t^\prime)^2\right )
-{{g}\over{12}}(t^{\prime\prime}-t^\prime)^3\right ]
-{{(x^\prime)^2}\over{2\sigma^2}}}\int_{-\infty}^{+\infty}dye^{Ay^2+By} 
\eeaq
where 
\beq
A=-{{1}\over{2\sigma^2}}+{{im}\over{2\pi (t^{\prime\prime}-t^\prime)}} 
\ \ \ and \ \ \ B={{x^\prime}\over{\sigma^2}}+i({{-m}
\over{2\hbar (t^{\prime\prime}-t^\prime)}}(2x^{\prime\prime}
+g(t^{\prime\prime}-t^\prime)^2)+k_0). 
\eeq
\ewt
Using the analytic continuation of the integral
\beq 
\int_{-\infty}^{+\infty}dye^{-ay^2+iby}
=\sqrt{{{\pi}\over{a}}}e^{-{{b^2}\over{4a}}}, \ \ \ a>0, \ \ \ b\in \R 
\eeq
to the domain $a\in \C$, $Re(a)>0$, $b\in \C$, we obtain
\beq
\psi(x^{\prime\prime},\dt,x^\prime;g;\sigma;
{{m}\over{\hbar}};k_0)=\sqrt{{{\sigma}
\over{i\sqrt{\pi}\dt {{\hbar}\over{m}}
+\sqrt{\pi}\sigma^2}}}
e^{\cal F} e^{\cal G}, 
\eeq
where for convenience we have defined
\beq
\dt=t^{\prime\prime}-{t^\prime},
\eeq
\bwt
\beaq
{\cal F}&=&
-\frac{1}{2} \left [ \frac{{x^\prime}^2}{\sigma^2}
+
\left ( 
\left \{
- {x^\prime}^2 \dtt + 
\frac{m^2}{4\hbar^2}\sigma^4 (2x^{\prime\prime}+g\dtt)^2 
\right.\right.\right.
\nonumber\\
&&\left. \left. \left.
+ k_0\sigma^4\dt [k_0\dt-\frac{m}{\hbar} (2x^{\prime\prime}+g\dtt)]\right \}
\sigma^2 \dtt
\right.\right.
\nonumber\\
&&\left .\left.
-x^{\prime}\frac{m^2}{\hbar^2}\sigma^6\dtt
(2x^{\prime\prime}+g\dtt)
+\frac{2m}{\hbar}\sigma^6k_0x^{\prime}{(\dt)}^3\right )
[
\sigma^4{(\dt)}^4 (1+\frac{m^2\sigma^4}{\hbar^2 \dtt})
]^{-1}
\right ],
\eeaq
and
\beaq
{\cal G}&=&
-\frac{i}{2}\left [
x^{\prime}{(\dt)}^3\frac{m}{\hbar}\sigma^4(2x^{\prime\prime}+g\dtt)
-\frac{m}{\hbar}\sigma^4 {x^\prime}^2{(\dt)}^3
+\frac{m^3}{4\hbar^3}\sigma^8\dt (2x^{\prime\prime}+g\dtt)^2
\right.\nonumber\\
&&\left.+\left (
k^2_0\sigma^4\dtt-\frac{m}{\hbar}\sigma^4 k_0\dt(2x^{\prime\prime}+g\dtt)
\right) \frac{m}{\hbar}\sigma^4\dt
\right.\nonumber\\
&&\left.
-2\sigma^4 k_0 x^{\prime}{(\dt)}^4 \right ]
[\sigma^4{(\dt)}^4 (1+\frac{m^2\sigma^4}{\hbar^2 \dtt})
]^{-1}.
\eeaq
\ewt
The factor before the exponentials has a well defined classical limit since 
\beq
\sqrt{{{\sigma}\over{i\sqrt{\pi}(t^{\prime\prime}-t^\prime){{\hbar}
\over{m}}+\sqrt{\pi}\sigma^2}}}\buildrel {\hbar \to 0}\over 
\longrightarrow {{\pi^{-{{1}\over{4}}}\over{\sqrt{\sigma}}}}; 
\eeq
however, the imaginary exponential does not exist in this limit since it
 behaves as 
\beq
exp
\left [
-{{i}\over{2}}{{m}\over{\hbar}}{{
\left (x^{\prime\prime}
+{{g}\over{2}}(t^{\prime\prime}-t^\prime)^2\right )^2}
\over{t^{\prime\prime}-t^\prime}}+i{{p_0}\over{\hbar}}
\left (x^{\prime\prime}+{{g}\over{2}}(t^{\prime\prime}-t^\prime)^2
\right )\right ]
\eeq
which oscillates indefinitely in the unit circle when $\hbar \to 0$ 
(unless $u_0={{1}\over{2}}{{x^{\prime\prime}+{{g}
\over{2}}(t^{\prime\prime}-t^\prime)^2}\over{t^{\prime\prime}-t^\prime}}$, 
in which case is equal to 1). 
This means that the wave function does not exist in this limit. However, 
this is not so for the classical limit of the probability density.
After some reordering of the terms in the real exponential, the square of the 
absolute value of the wave function, namely, the probability density 
($\rho$) is given by 
\bwt
\beaq
&&\rho(x^{\prime\prime},t^{\prime\prime}-t^\prime,x^\prime; 
g;\sigma;\mu;u_0)
=\vert\psi(x^{\prime\prime},t^{\prime\prime}
-t^\prime,x^\prime; g;\sigma,\mu;u_0)\vert ^2
\nonumber\\
&&={{1}\over{\sqrt{\pi}\sqrt{{{(t^{\prime\prime}
-t^\prime)^2}\over{\mu^2\sigma^2}}+\sigma^2}}}
~ exp
\left [-{{\left (x^{\prime\prime}-
\left [x^\prime+u_0(t^{\prime\prime}-t^\prime)
-{{g}\over{2}}(t^{\prime\prime}-t^\prime)^2\right ]
\right )^2}\over{{(t^{\prime\prime}
-t^\prime)^2}\over{\mu^2\sigma^2}}+\sigma^2}\right ],
\label{probdensity} 
\eeaq
\ewt
where we have defined $\mu=m/\hbar$. Clearly,
\beq 
\int_{-\infty}^{+\infty}dx^{\prime\prime}\rho(x^{\prime\prime},
t^{\prime\prime}-t^\prime,x^\prime;g;\sigma;\mu;u_0)=1,
\eeq 
i.e. the normalization of the initial wave function is preserved. Also,
\beq 
\rho(x^{\prime\prime},t^{\prime\prime}-t^\prime,x^\prime;g;\sigma;\mu;u_0)
=\rho(x^{\prime\prime}-x^\prime,t^{\prime\prime}-t^\prime;g;\sigma;\mu;u_0).  
\eeq
Eq.(\ref{probdensity}) illustrates the {\it violation of the WEP}: 
the probability 
density to find the ``falling" quantum particle  at $x^{\prime\prime}$ and at 
time $t^{\prime\prime}$ depends on the mass $m$ through the ratio $\mu$, 
showing that quantum non-relativistic particles ``fall" differently for 
different values of the mass. In the next section we show, however, that 
the WEP is recovered in the limit $\hbar \to 0$. 
The width $\Sigma$ of the probability density $\rho$ decreases with the mass 
and is given by 
\beq
\Sigma=\sigma
\sqrt{ 1+ \frac{\hbar^2 (t^{\prime\prime}-t^{\prime})^2}{m^2\sigma^4}}.
\eeq
Notice that $\Sigma$ is independent of $g$ and coincides with the broadening 
of a free gaussian wave  packet. 
Since $\rho$ is normalized, for $m_1>m_2$, $\rho_1$ is more peaked 
than $\rho_2$ and so $\rho_1(x^{\prime\prime})>\rho_2(x^{\prime\prime})$ 
for $\vert x^{\prime\prime}-\bar{x^{\prime\prime}}\vert <\Delta$ and 
$\rho_1(x^{\prime\prime})<\rho_2(x^{\prime\prime})$ for 
$\vert x^{\prime\prime}-\bar{x^{\prime\prime}}\vert >\Delta$ with 
\beq
\Delta={{\Sigma_1\Sigma_2}\over{\sqrt{\Sigma_2^2-\Sigma_1^2}}}
\sqrt{ln({{\Sigma_2}\over{\Sigma_1}})}, 
\eeq
defined by
$\rho_1(\bar{x^{\prime\prime}}\pm \Delta)
=\rho_2(\bar{x^{\prime\prime}}\pm \Delta)$.
Here, 
 \beq
\bar{x^{\prime\prime}}=x^\prime+u_0(t^{\prime\prime}
-t^\prime)-{{g}\over{2}}(t^{\prime\prime}-t^\prime)^2. 
\eeq
Far away from the center of the distributions, the lighter particle 
has more probability to be found i.e. to have fallen, than the heavier one.

In terms of $\bar{x^{\prime\prime}}$
and $\Sigma$, the probability density at 
$x^{\prime\prime}$ is given by 
\beq
\rho(x^{\prime\prime};\bar{x^{\prime\prime}},\Sigma)
={{1}\over{\sqrt{\pi}\Sigma}}e^{-{{(x^{\prime\prime}
-\bar{x^{\prime\prime}})^2}\over{\Sigma^2}}}. 
\eeq
In particular, when the average velocity $u_0$ of the initial wave packet 
equals the initial velocity $v_0$ of the classical solution, 
$\bar{x^{\prime\prime}}=x^{\prime\prime}$ and $\rho$ reaches its maximum 
value: 
\beq
\rho(x^{\prime\prime};x^{\prime\prime},\Sigma)
=\rho_{max}={\frac{1}{\sqrt{\pi}\Sigma}} \ \ for \ \ u_0=v_0. 
\eeq
For a qualitative picture of the above mentioned behavior of quantum falling,
in Fig. 1, we plot $\rho$ as a function of $x^{\prime\prime}$ ( in
units of meters) for $\pi^0$ (mass 134.98 $MeV/c^2$) and $\pi^{\pm}$ (mass
139.57 $MeV/c^2$); in Fig. 2 , we plot $\rho$ for $\pi^0$ and  $K^0$ ( mass
497.67 $MeV/c^2$). In both cases we have chosen ${x^{\prime}}=0$ and 
$u_0$ such that ${\bar{x^{\prime\prime}}}=0$. We have also taken 
$\sigma=10^2~ \AA$.  In both figures, it is clear that far away from the 
center of the distributions, the lighter particle
has higher probability to be found than the heavier one.


\begin{figure}[t!] 
\vspace{0.5cm}
{\centering
\resizebox*{0.4\textwidth}{0.2\textheight}
{\includegraphics{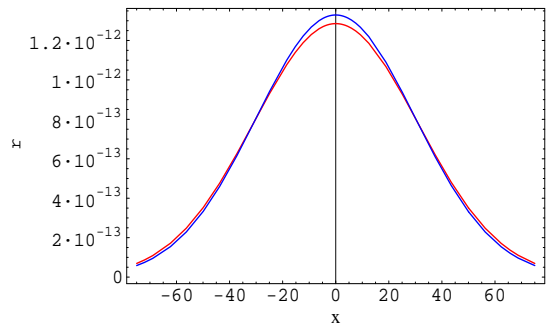}}
\par}
\caption{
The probability density $\rho\equiv r$ is plotted as a function of 
${x^{\prime\prime}}\equiv x$ for $\pi^{\pm}$ (upper curve at x=0)  
and $\pi^{0}$.}
\label{fig1}
\end{figure}

\begin{figure}[t!] 
\vspace{0.5cm}
{\centering
\resizebox*{0.4\textwidth}{0.2\textheight}
{\includegraphics{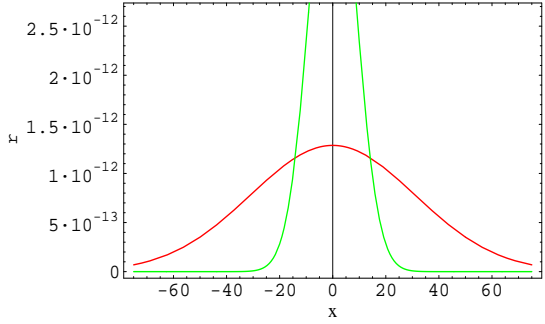}}
\par}
\caption{
The probability density $\rho\equiv r$ is plotted as a function of 
${x^{\prime\prime}}\equiv x$ for $\pi^0$ and $K^0$ (upper curve at x=0).}
\label{fig2}
\end{figure}


\section{Classical limit and recovery of the WEP}

The Eq.(\ref{probdensity}) for the probability density at 
$(x^{\prime\prime},t^{\prime\prime})$ for the freely falling quantum particle 
in the constant and uniform gravitational field $\vec{g}=-g\hat{x}$, has a 
well defined {\it mass independent} classical limit given by
\bwt
\beq 
lim_{\hbar \to 0}\rho(x^{\prime\prime}-x^\prime,t^{\prime\prime}
-t^\prime;g;\sigma;{{m}\over{\hbar}};u_0)
={{1}\over{\sqrt{\pi}\sigma}}exp-{{(x^{\prime\prime}
-\bar{x^{\prime\prime}})^2}\over{\sigma^2}}\equiv\rho_{cl}(x^{\prime\prime}
-x^\prime,t^{\prime\prime}-t^\prime;g;\sigma;u_0).
\label{limhbar0} 
\eeq
\ewt
The gaussian in Eq.(\ref{limhbar0}) has the same width as the initial 
probability 
distribution, but is centered around $\bar{x^{\prime\prime}}$. 
The absence of the mass in $\rho_{cl}$ exhibits the recovery of the WEP in 
the classical limit\cite{davies}. 
For the case of an initial perfectly localized particle, with probability 
density 
\beq
lim_{\sigma \to 0}\vert\psi(y,t^\prime)\vert^2
={{1}\over{\sqrt{\pi}}}lim_{\sigma \to 0}{{e^{{(y-x^\prime)^2}
\over{\sigma^2}}}\over{\sigma}}
=\delta(y-x^\prime), 
\eeq
one also obtains a perfectly localized particle in the classical limit: 
\bwt
\beaq
&&lim_{\sigma \to 0}\rho_{cl}(x^{\prime\prime}-x^\prime, t^{\prime\prime}
-t^\prime;g;\sigma;u_0)
={{1}\over{\sqrt{\pi}}}lim_{\sigma \to 0}{{e^{{(x^{\prime\prime}
-\bar{x^{\prime\prime}})^2}\over{\sigma^2}}}\over{\sigma}}
\nonumber\\
&&=\delta(x^{\prime\prime}-\bar{x^{\prime\prime}})
=\delta
\left (x^{\prime\prime}-\left [x^\prime +u_0(t^{\prime\prime}-t^\prime) 
-{{g}\over{2}}(t^{\prime\prime}-t^\prime)^2\right ]\right ).
\label{prodencl}
\eeaq
\ewt
So our probability density in Eq.(\ref{prodencl}) is independent of mass and
localized in space, which is consistent with the classical description of the 
particle.

\begin{center}
{\bf ACKNOWLEDGMENTS}
\end{center}
This work is partially supported by the grants PAPIIT-UNAM 
IN103505 (M.S., S.H.) and 
IN94045 (S.S.). M. S.  thanks the Abdus Salam International Centre of 
Theoretical 
Physics, Trieste, for support through the Net 35, and Augusto Gonz\'alez of 
the ICIMAF, La Habana, Cuba, for an enlightened discussion.

\end{document}